\documentclass{article}

\pdfoutput=1

\usepackage{arxiv}

\usepackage[utf8]{inputenc} 
\usepackage[T1]{fontenc}    
\usepackage{hyperref}       
\usepackage{url}            
\usepackage{booktabs}       
\usepackage{amsfonts}       
\usepackage{nicefrac}       
\usepackage{microtype}      
\usepackage{lipsum}
\usepackage{graphicx}
\graphicspath{ {./images/} }

\title{Implicit Gender Bias in Computer Science---A~Qualitative Study}

\author{
 Aurélie Breidenbach \\
  Department for Political Sciences and Sociology\\ 
  University of Bonn, Faculty of Arts\\
  53113 Bonn, Germany \\
  \texttt{aurelie.breidenbach@outlook.com} \\
   \And
 Caroline Mahlow \\
  Institute for Software Technology\\
  German Aerospace Center (DLR)\\
  51147 Cologne, Germany \\
  \texttt{caroline.mahlow@dlr.de} \\
  \And
 Andreas Schreiber \\
  Institute for Software Technology\\
  German Aerospace Center (DLR)\\
  51147 Cologne, Germany \\
  \texttt{andreas.schreiber@dlr.de} \\
}

\begin{document}
\maketitle
\begin{abstract}
Gender diversity in the tech sector is---not yet?--sufficient to create a balanced ratio of men and women. For many women, access to computer science is hampered by socialization-related, social, cultural and structural obstacles. The so-called implicit gender bias has a great influence in this respect. The lack of contact in areas of computer science makes it difficult to develop or expand potential interests. Female role models as well as more transparency of the job description should help women to promote their---possible---interest in the job description. However, gender diversity can also be promoted and fostered through adapted measures by leaders.
\end{abstract}


\section{Introduction}

The concept of \emph{implicit gender bias} describes cognitive bias effects that influence a person's personal perception, actions and behavior. In contrast to an individual's explicit and intentional actions, implicit gender bias corresponds, for example, to unconscious assumptions, attitudes, stereotypes and prejudices about individuals or groups of individuals, in this case about women. This manifests itself in discrimination, sexism, gender stereotyping and causes structural inequalities.

Looking back to the beginnings of computer science, women were significantly involved in writing the first codes. In the basic research of this field, it was possible to work without prejudice or bias. A change in gender diversity took place in the 1980s, when the number of home computers increased rapidly. Boys were encouraged more by their parents than girls, thus establishing the view that technology and computers were an area for men. Until now, the proportion of women in information technology professions is far below that of men. Such gender stereotypes influence how the skills of men and women are perceived and how their performance is interpreted.

All individuals have implicit attitudes and presuppositions towards people. In everyday life, categorizations facilitate unconscious classification into certain ``drawers''~\cite{HosodaStoneRomeroCoats2003}. Different expectations arise about their temperament, their behavior, their abilities and skills as well as their choice of occupation.

Most of the existing software engineering literature focuses on explicit gender bias but ignores implicit gender biases~\cite{WangRedmiles2019}. Consequently, the question arises whether implicit gender biases in the tech sector have an impact on the number of women in IT\@. Can the female gender be implicitly influenced?

A further explanation for the low rate of women in IT is the biological approach, which describes the lack of analytical skills among women as a cause. This and other differences between men and women that could explain possible causes of the lack of diversity were presented by James Damore in 2017~\cite{Damore2017}.

The then 23-year-old software developer at Google recorded prejudices and stereotypes about women in an internal memo, with which he wanted to draw attention to the diversity issues in Google's corporate culture.

Damore argues that gender differences are not socially constructed but biologically determined. This is due to the prenatal testosterone that unborn children receive in the womb. This would give boys and men an advantage in analytical and mathematical thinking, which would give a biological advantage in software development.

Many of the differences are marginal and there are several and significant overlap between the sexes, so they have nothing on individuals would testify~\cite{Damore2017}. However, it does address concrete differences in the behavior and preferences of men and women. For example, women in the IT sector are more likely to work in front-end development, have a social disposition and prefer to work in a communicative way. Men, on the other hand, prefer abstract and analytical work in back-end software development and are more interested in job-related ``status'' and high pay. Therefore, women with social advantages could be lured into the tech industry by adjusting conditions such as a better ``work-life balance''.

However, Damore sees no need to create a gender balance within the tech industry. People should rather learn to deal with ``natural'' differences by training their empathy and paying attention to biological differences and causes~\cite{Damore2017}. Furthermore, he refers to the effect of institutional culture through management and the unconscious behavior of employees that goes together with it.

\section{Qualitative Study}
\label{sec:qualitative_study}

To verify Damore's statements, we investigate which determinants of the implicit gender bias women encounter in their lives and which factors influence or have influenced the career choices of women in computer science. In addition, this connection is linked to the aspect of leadership and corporate culture and the influence of the supervisor.

\begin{figure}[tb]
  \centering
  \includegraphics[width=\textwidth]{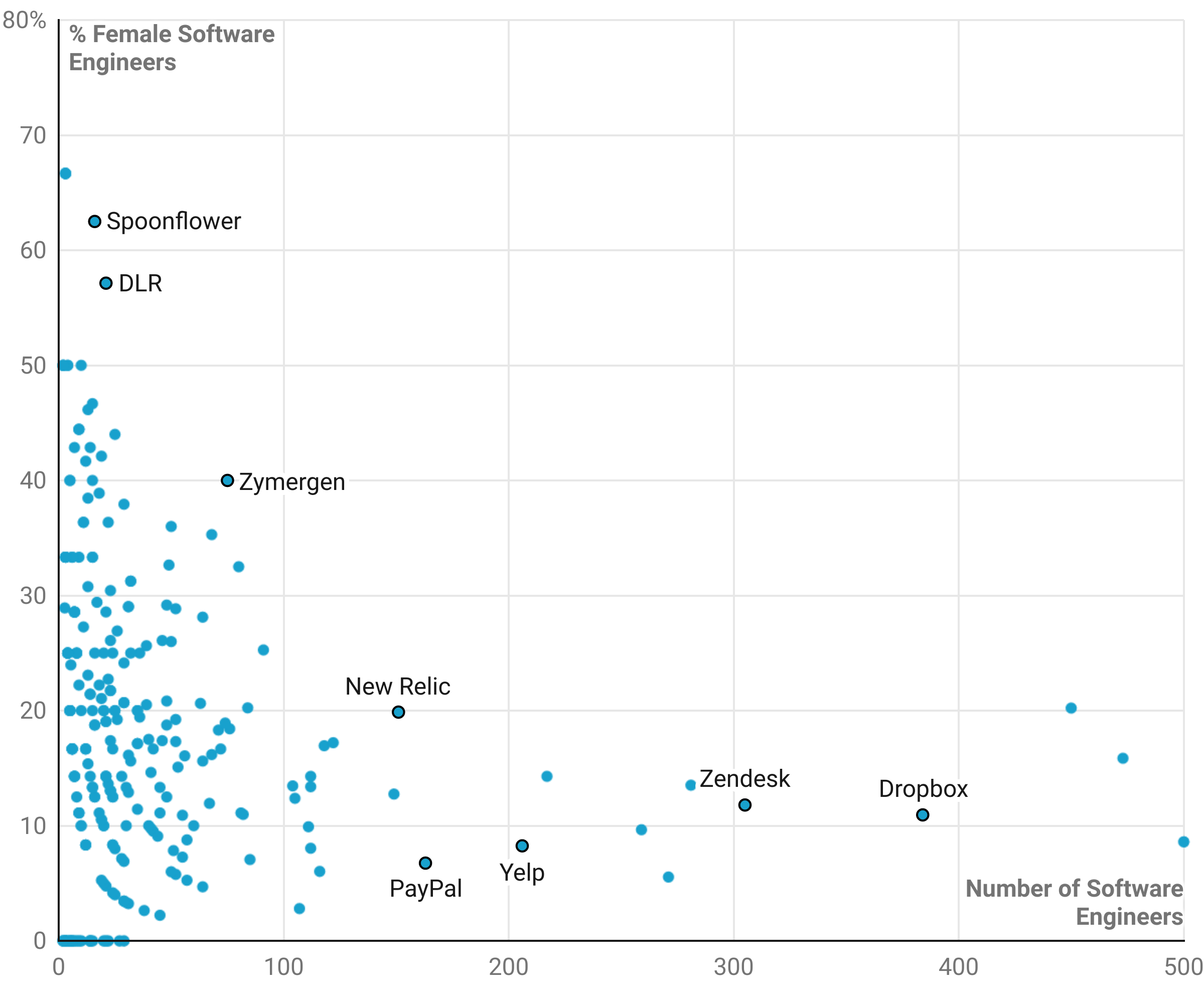}
  \caption{Women in software engineering: percentage of women in software engineering teams (permanently employed staff only). URL: \url{https://datawrapper.dwcdn.net/qPx96/2/}}
  \label{fig:women-in-software-engineering}
\end{figure}

We conducted an (unrepresentative) empirical qualitative study on gender bias in computer science~\cite{Breidenbach2020} using the example of the staff of the Institute for Software Technology of the German Aerospace Center (DLR). This institute has an unusually high proportion of women (see ``DLR'' in Figure~\ref{fig:women-in-software-engineering}) and allows us to question whether corporate culture has an influence.

Five women from the Intelligent and Distributed Systems Department were interviewed. Three of them were researchers, one student and one team. This cross-section of professional degrees gives an indication which determinants are decisive for women in their choice of career and what experiences they have had in their previous lives. 

To ensure a comparison between the sexes and to find out whether and which measures are used to make the corporate culture more pleasant for women and to establish a positive working climate for all employees, the head of department was also interviewed as a managerial position. Through the long period of employment at DLR ($>21$ years), the latter is able to describe the long-term trend in corporate structure and culture.

Public opinion on Damore's theses ranges from strong support to strong rejection, which is why the memo provides a good basis for a key stimulus in the qualitative survey. All interview partners receive the key stimulus to read independently. The key stimulus is followed by four statements on which the interviewees are asked to comment. These represent the essential theses of Damore.

The actual interview begins, divided into three levels:

\begin{enumerate}
    \item The first level resembles a biographical interview, a narrative way of life and career relating to IT experiences. Subsequently, further determinants of gender prejudices in the professional field of computer science are asked. 
    
    \item The second level revels expectations and preferences towards the employer and examines Damore's statements about gender preferences in the decision to choose a profession. 
    
    \item The third level is dedicated to the personal experiences of women regarding prejudices, and discrimination which women face in the professional world and their personal careers.
\end{enumerate}

The questionnaire of the management position differs from that of women in that is more focused on the aspect of corporate management and culture.

To cover as many sub-areas of the subjects' experiences as possible, the type of questions was adapted to the interview and varied between open (question about biography), semi-open (explicit question about experiences), and closed questions (has something already been experienced by the women).

\section{Results}
\label{sec:results}

Overall, the subjective views of the respondents prove that the tech sector is still a male domain. 

According to the interviewees, this representation has become entrenched in society through implicit gender bias rather than through cognitive and physical characteristics of men. The first indications of the influence of implicit gender bias on the male domain in the tech sector are provided by the survey of the autobiographical life and career path of the interviewees.

\subsection{The Way Into the IT Industry}
\label{sec:the_way_into_the_it_industry}

The first IT experiences were already made in childhood. The contact persons were mostly male relatives. 

By interacting with computers in childhood, the women had no fear of contact with the devices or programs. This made it easier for them to further access to the subject of computer science and increased interest. This developing interest could be continued during the school years by taking compulsory subjects or at least optional subjects with computer science background. Only one respondent described that no experience with computers was gained at school. This also had an effect on the choice of study. She studied mathematics, since computer science was never present. 

The women reported conversations with other women who described computer science not to be able to study because before the beginning of their studies, they were lacking programming skills. This seems to be common among young women, who are deterred and unsettled by it. This suggests that women tend to underestimate themselves and often do not dare to tackle topics with which they have not yet had any contact. The prevailing image of the computer scientist by society creates an additional barrier, so that access to this subject becomes difficult or even impossible as a career option.

All respondents confirmed that the proportion of women during their studies was very low. Even though women were not discriminated against by professors and lecturers directly, they did encounter some statements that also suggest the implicit gender bias. These mostly came from (male) fellow students in the form of ignorance or a false assessment of their skills and knowledge. 

Despite the initial difficulties, all respondents were very confident in their choice and decided to continue their career path in the tech industry. The many job opportunities in the IT industry and the lack of skilled workers argued for a job with a future.

Such gender-related distortions remain in professional life. For example, one respondent described a situation at a trade fair where she had to justify herself several times that she was doing software development, since she ``\emph{obviously has nothing to do with Python.}'' This phenomenon intensifies the more feminine and cultivated she is in a male-dominated environment. Not only men would have this prejudice against women, but especially women against women. In contrast, she also reports that she has had the experience that men have to justify themselves at a conference if they have nothing to do with programming.

The head of department also reported on some implicit gender bias that he had experienced over time and collected from colleagues and acquaintances. These almost exclusively proved to be a disadvantage for women, for example, in that women were overheard or not taken seriously in discussions. 

This is also confirmed by the women interviewed. One of the interviewees was told that she would positively influence the mood and make it more pleasant. Overall a positive working environment was described in which one would be treated equally and fairly. The high proportion of women has a positive influence on social behavior and promotes teamwork.

It should also be emphasized that being a woman in a male domain does not only have negative effects. The interviewees reported a kind of unique selling proposition, which could have a positive effect on the recognition value.

\subsection{Decision Criteria for the Tech Sector}
\label{sec:decision_criteria_for_the_tech_sector}

When choosing an employer, all respondents seem to share the same basic values and premises. All women stated that they wanted to pursue a profession that would fulfill and challenge them. The work should be fun and the topic should be interesting. The women thus decided to choose DLR as their employer because of the variety of topics and their interest in DLR’s research fields. 

The aspect of work-life balance only becomes more and more important for the women computer scientists as their length of employment increases. In job interviews, many women are relieved when the topic of family planning would be openly addressed and treated as a matter of course.

Monetary and other social advantages and possibilities of work allocation are a bonus for the women, but these were not decisive criteria in their application. More important to women is a good working environment in which they are taken seriously, and all colleagues are treated equally.

\subsection{Corporate Culture}
\label{sec:corporate_culture}

The department promotes active leadership by women, dissolve ``typical'' role models, and initiate gender-neutral actions. This begins with the most neutral formulation of job advertisements. As soon as an employee notices a case that can be described as sexist, misogynist, or exclusionary, an appraisal interview is arranged. The aim is to create awareness and to reflect on one's own behavior and to work out alternatives.

In addition, the department leader must set an example of gender-neutral behavior and trust. Thus, the priorities of health and family are equally important to all employees. In this respect, trust also means trusting in the abilities and freedom of employees by working independently.

For the application process, a procedure was developed which every candidate goes through when applying. In the decision-making process, every vote of potential team members and colleagues counts—including a strict ``veto'' right for everybody. This procedure is intended to ensure that no personnel decisions are based on individual opinions. In addition, the focus of professional strengths was shifted to soft skills, while in the past applicants were $100\%$ matched by pure technical skills. Due to the changed focus, all noticed better interpersonal vibes within teams—and more positive group dynamics also enhances the productivity of the department.

During the application process it was often found that mentioning topics such as hobbies relieves the tension of the candidates. Furthermore, it is helpful to mention in job interviews how the corporate culture and values are lived in the department.

\section{Conclusion and Future Directions}
\label{sec:conclusion}

The following discussion is based on the results of the interviewed people from the software industry of a common department. Therefore, from the subjective experiences and reports are not generally applicable to other persons or companies will be closed. Nevertheless, these subjective experiences form a real state.

All statements of Damore are discussed on a biological level. Although he is trying several times, his statements regarding the differences between men and women are not absolute, yet they are natural and not culturally influenceable. The fact that computer science, especially development of innovative software solutions, once was a female industry, contradicts this concept.

The fact that women appreciate the working environment and the social benefits is in line with Damore's statements, but this fact does not provide the motivation for applying for a job with a company. 

Rather, it shows that cognitive gender bias is related to career choice and thus influences gender diversity within the tech industry. The origin of the determinants of implicit gender bias in the occupational field of computer science lies in childhood. Depending on the support provided, certain skills develop earlier than others. Capabilities and skills therefore differ between individuals and not between genders. From this it can be deduced that one of the obstacles to the lack of women in IT is not the lack of natural interest and cognitive skills, as Damore puts it, but rather the lack of contact in the CV and cognitive barriers caused by gender bias.

The connection between the external appearance and the job computer scientist also illustrates the extent to which stereotypes are anchored in the IT industry and the understanding of social roles is shaped and promoted by culture. This insight thus contradicts Damore's thesis that gender differences are not socially constructed. 

The resulting stereotypes, prejudices and cognitive distortions lead to unconscious evaluations and micro-behaviors towards the female, but also towards the male gender. The classic role models still exist but seem to be slowly dissolving—with a positive effect on group dynamics, the variety of topics, and productivity.  This experience supports the positive effect of gender balance in software teams and—in contrast to Damore's renunciation of gender diversity—speaks for more women in the software industry.

Therefore, a value balance between the sexes should be promoted so that people can pursue the profession they are interested in without being judged too quickly. Transparency, an awareness of these distorting effects, and setting an example of equal rights through management positions can contribute to this.

\bibliographystyle{unsrt}  
\bibliography{arxiv-implicit-gender-bias-in-computer-science}  

\end{document}